\documentclass[conference]{IEEEtran}
\IEEEoverridecommandlockouts

\usepackage{cite}
\usepackage{url}
\usepackage{amsmath,amssymb,amsfonts}
\usepackage{algorithmic}
\usepackage{graphicx}
\usepackage{textcomp}
\usepackage{xcolor}
\usepackage{float}
\usepackage{booktabs}  % 用于三线表 (\toprule, \midrule, \bottomrule)
\usepackage{tabularx}  % 用于自动调整列宽
\usepackage{multirow}  % (可选) 如果需要跨行合并

\def\BibTeX{{\rm B\kern-.05em{\sc i\kern-.025em b}\kern-.08em
    T\kern-.1667em\lower.7ex\hbox{E}\kern-.125emX}}
\begin{document}

\title{DACP: A Scientific Data Access and Collaboration Protocol\\
}

% \and
% \and
% \and
% \and
% \and

\author{\IEEEauthorblockN{Zhihong Shen$^{1,2,*}$, Xiaojie Zhu$^{1,2}$, Zhenjing Cheng$^{1}$, Hao Ren$^{1}$, Zhaoji Liang$^{1,2}$, Changfa Lu$^{1}$}
\IEEEauthorblockA{\textit{Computer Network Information Center, Chinese Academy of Sciences$^1$}, Beijing, China \\
\textit{University of Chinese Academy of Sciences$^2$}, Beijing, China \\
\{bluejoe, xjzhu, zjcheng, rh, zjliang, luchangfa\}@cnic.cn}
}

\maketitle

\begin{abstract}
Scientific computing is rapidly entering a data-intensive era. However, existing general-purpose network protocol stacks face limitations in eliminating data silos and improving data accessibility and interoperability, making it difficult to effectively meet the demands of emerging paradigms such as AI4Science. To address these challenges, we propose the Data Access and Collaboration Protocol (DACP). DACP defines the Streaming Data Frame (SDF) as its core data model. Through Unified Resource Identification, columnar stream framing, and a reverse supply mechanism, DACP enables data discovery, in-situ computation, and the streaming return of results across scientific data centers, thereby facilitating efficient cross-domain collaboration. Furthermore, this paper introduces faird, a reference server implementation of DACP. This work provides a viable path for building scalable and collaborative scientific data infrastructures.
\end{abstract}


\section{Introduction}
In recent years, with the deepening of scientific inquiry, scientific computing is rapidly evolving toward a data-centric paradigm\cite{hey2009fourth}. Domains such as astronomical observation, high-energy physics, and life sciences are continuously generating data at the Petabyte (PB) or even Exabyte (EB) scales. As data becomes a core infrastructure, the capability to acquire, comprehend, and utilize it directly determines the efficiency and boundaries of scientific discovery.

Currently, the FAIR principles (Findable, Accessible, Interoperable, Reusable)\cite{wilkinson2016fair} have become a global consensus for scientific data management. However, current FAIR practices focus primarily on ``whether data exists and is describable''. Particularly in cross-disciplinary, cross-institutional, and cross-data-center scenarios, data accessibility and interoperability still have significant room for improvement.

Emerging research paradigms like AI4Science increasingly depend on joint training and inference over large-scale, multi-modal data\cite{wang2023scientific}. However, in practice, scientific data remains dispersed across fragmented systems characterized by heterogeneous formats, interfaces, and permission models, leading to severe data silos. This landscape of dispersion, heterogeneity, and siloization \cite{halevy2006data}\cite{shen2024} has emerged as a critical bottleneck hindering the advancement of these new paradigms.

\subsection{Problem Statement}\label{AA}
Although a variety of general-purpose data transfer and access protocols have been proposed and deployed, their direct reuse for scientific data still presents obvious shortcomings:

First, the lack of a unified abstraction exacerbates the problem of ``Data Silo''\cite{chen2014datasilo}. Different data sources are usually exposed as independent service interfaces with inconsistent access methods and naming rules. This leads to the need for extensive customized development at the application layer to achieve cross-source data composition.

Second, legacy file-centric models incur severe ``Read Amplification''\cite{abernathey2021cloud}. Researchers often require only partial fields, columns, or sub-regions of a dataset. However, existing protocols usually require downloading the complete file, bringing significant network and storage overheads.

Third, data and metadata are fragmented in the access path. Information such as data types, structures, and semantics cannot be transmitted synchronously with the data stream, increasing the complexity of cross-system interoperability.

Finally, existing protocols mainly focus on ``how to transmit data'' while ignoring ``where and how data should be processed''. In large-scale data scenarios, data must be fully downloaded to compute nodes, making it difficult to support in-situ computing\cite{balasubramonian2014near} and task-driven data supply.

\begin{table*}[htbp]
  \centering
  \caption{Comparison of Computer Network, WWW, Linked Data, and Research Data Network (Adapted from \cite{shen2024})}
  \label{tab:network_comparison}
  % 定义列格式：
  % l: 左对齐 (Left)
  % X: 自动换行 (Auto-wrap based on width)
  % c: 居中 (Center)
  \begin{tabularx}{\linewidth}{l X l X}
    \toprule
    \textbf{Network Type} & \textbf{Primary Objective} & \textbf{Core Element} & \textbf{Core Technology} \\
    \midrule
    Computer Network\cite{http1974} & 
    Shield heterogeneity of computing devices; provide interoperability between devices & 
    Devices & 
    IP, TCP, etc. \\
    \addlinespace % 增加行间距，使表格更易读
    World Wide Web (WWW)\cite{www1994} & 
    Shield heterogeneity of documents; provide interoperability between documents & 
    Documents & 
    URL, HTTP, HTML, Hyperlink, etc. \\
    \addlinespace
    Linked Data\cite{semantic2006} & 
    Shield heterogeneity of data (information); provide interoperability between data & 
    Resource Description & 
    URI, HTTP, RDF, RDF link, etc. \\
    \addlinespace
    \textbf{Research Data Network}\cite{shen2024} & 
    Shield heterogeneity of research data; provide interoperability between data and services & 
    Research Data & 
    Identification, Addressing, Access, and Operation Protocols for research data resources \\
    \bottomrule
  \end{tabularx}
\end{table*}

\subsection{Solution Approach and Contributions}\label{AA}
To address the aforementioned challenges, the main contributions of this paper are as follows:

\begin{itemize}
\item We proposed DACP, a scientific data access and collaboration protocol. This approach provides a unified, task-oriented, and computation-centric abstraction for scientific data access, enabling efficient data collaboration across Wide Area Networks (WANs).

\item We presented faird, a scalable reference implementation based on the DACP protocol.The source code is publicly available at \url{https://github.com/rdcn-link/faird}.
\end{itemize}

The remainder of this paper is organized as follows: Section II analyzes existing limitations. Sections III and IV detail the DACP design and faird implementation. Section V evaluates the performance, followed by application scenarios in Section VI and conclusions in Section VII.

\section{Analysis of Limitations in Existing Protocol Stacks}

A Research Data Network\cite{shen2024} constitutes a unified and standardized infrastructure designed for distributed scientific data. As illustrated in Table 1, significant distinctions exist between research data networks and the World Wide Web (WWW) regarding their objectives, core elements, and underlying technologies. This section provides a critical analysis of typical protocol stacks.

\subsection{REST/JSON}\label{AA}
REST/JSON represents the most ubiquitous paradigm for data access and exchange \cite{fielding2002principled}. Consequently, it imposes significant limitations within the context of Research Data Networks:

\begin{itemize}
\item \textit{Serialization Bottlenecks}: Scientific data typically consists of dense numerical matrices. As a text-based protocol, JSON incurs substantial CPU overhead due to extensive serialization and deserialization operations\cite{maeda2012performance}.
\item \textit{Type System Deficiency}: The JSON type system supports only a generic Number type and lacks the capability to explicitly distinguish between critical scientific types such as int8, uint64, float32, and float16. 
\item \textit{Lack of Native Slicing}: Accessing high-dimensional data often requires extracting specific columns or sub-regions; however, standard REST APIs typically mandate the transmission of the complete JSON structure.
\end{itemize}

\begin{figure*}[htbp]
\centerline{\includegraphics[width=0.9\textwidth]{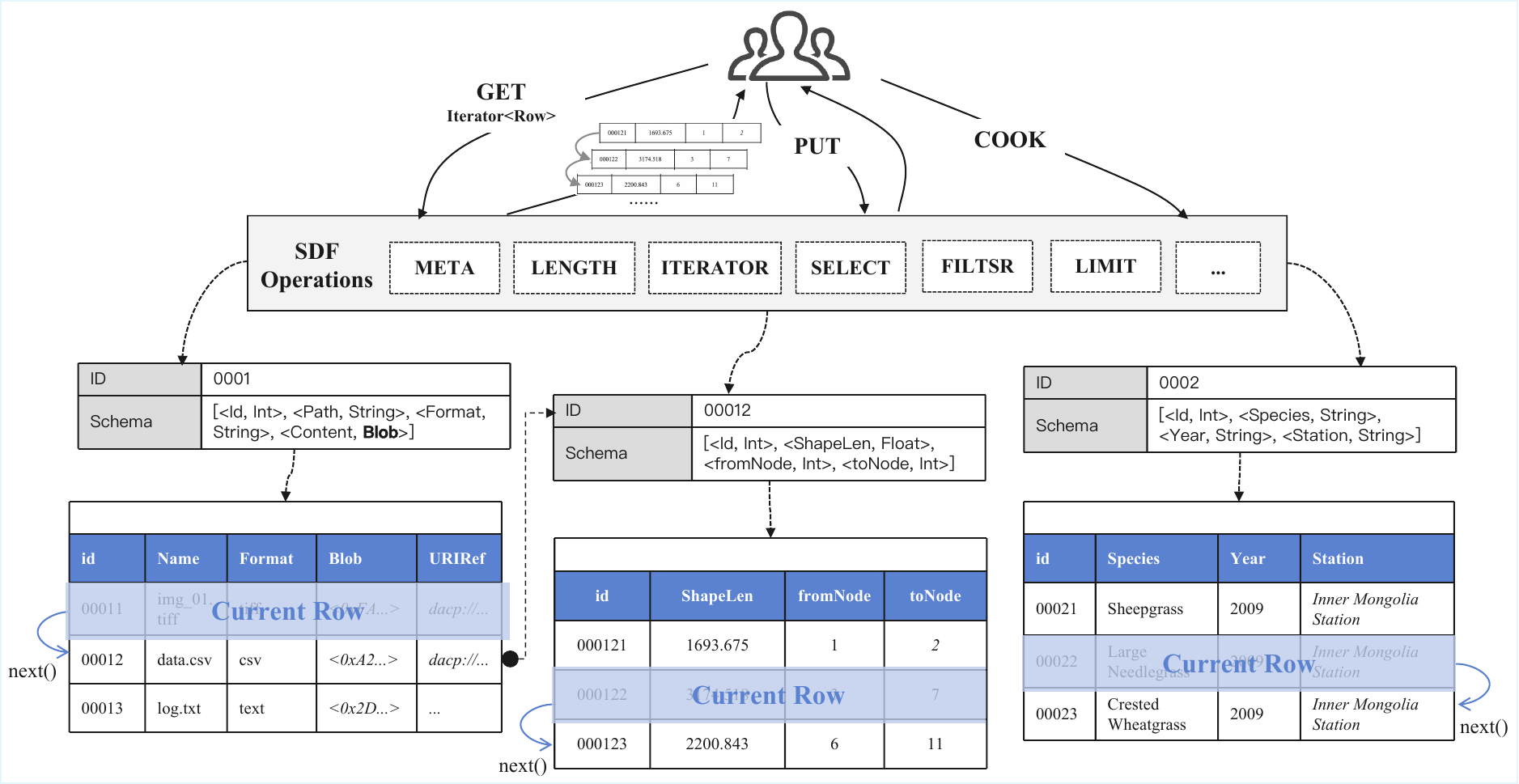}}
\caption{ \textbf{The unified logical abstraction and interaction architecture of the Streaming Data Frame (SDF)}. The figure illustrates the core protocol methods (GET/PUT/COOK), the streaming retrieval mechanism via GET, and the File List Framing strategy. Users can perform coarse-grained filtering on the file list (e.g., Format='csv') or recursively 'drill down' and retrieve specific file contents using the same set of operation primitives.
}
\label{dataframe}
\end{figure*}

\subsection{FTP/GridFTP}\label{AA}
FTP\cite{gien1978file} and GridFTP\cite{allcock2005globus} were primarily designed for high-throughput bulk data migration. Their architectures are inherently constrained by the file as the fundamental unit of transfer. Under emerging research paradigms characterized by online exploration and on-demand analysis, their limitations are becoming increasingly pronounced:

\begin{itemize}
\item \textit{Read Amplification}: Scientific tasks often require only specific columns, rows, or slices within a dataset. Yet, FTP transmission mandates the download of entire CSV or HDF5\cite{koranne2010hierarchical} files, which leads to the read amplification problem\cite{abernathey2021cloud}.
\item \textit{Opaque Internal Structure (Schema)}: The internal data structure (Schema) remains a ``black box'' to the transport layer. Users are compelled to download the file header or even the entire file before they can extract metadata.
\end{itemize}

In addition, other modern extension protocols, such as WebDAV\cite{dusseault2007webdav} and QUIC\cite{langley2017quic}, do not directly address the specific challenges of scientific data access. WebDAV remains fundamentally a file-centric protocol. In large-scale data center scenarios, the overhead associated with parsing its XML-based metadata often becomes a significant performance bottleneck. QUIC is strictly positioned at the Transport Layer. As such, it is agnostic to data models and does not concern itself with application-level logic regarding where or how data should be computed or composed.

In summary, current protocol stacks still exhibit certain deficiencies when applied to the accessibility and interoperability of scientific data. Scientists are compelled to spend a disproportionate amount of time locating, downloading, parsing, and organizing data, rather than focusing on the scientific problems themselves.

\section{DACP Protocol Design}

DACP is architected as an application-layer streaming protocol. Its objective is not to replace underlying transport technologies, but rather to build upon existing network infrastructures to empower scientific data with unified representation, programmable access, and cross-domain collaboration capabilities. This section elaborates on the design of DACP.

\subsection{Core Data Model: Streaming Data Frame (SDF)}

The data model of DACP is designed to abstract away the heterogeneity of underlying storage systems, providing a unified, computation-oriented abstraction.

DACP establishes the \textbf{Streaming Data Frame (SDF)} as its fundamental data primitive. The SDF abstracts scientific data content into a logical, streaming two-dimensional tabular structure, exposing a unified access interface. Formally, an SDF is represented as a tuple:

\begin{equation}
    \mathcal{D} = \langle \mathbb{S}, \mathcal{F} \rangle
\end{equation}

\subsubsection{Schema ($\mathbb{S}$)}
defines the topological structure and column type constraints of the 2D data frame:
\begin{equation}
    \mathbb{S} = \{ (attr_1, \tau_1), \dots, (attr_m, \tau_m) \}
\end{equation}
Here, $attr_i$ denotes the column name, and $\tau_i$ represents the data type (supporting various scientific primitives such as \texttt{int}, \texttt{float}, and \texttt{binary}). The existence of the Schema eliminates parsing ambiguities inherent in heterogeneous data, ensuring the self-describing nature of the SDF and end-to-end type safety.

\subsubsection{Frame Stream ($\mathcal{F}$)}
The data entity is materialized as an ordered streaming sequence, denoted as $\mathcal{F}$. Each element $\beta_k$ represents a Record Batch, containing a finite set of rows strictly conforming to the constraints of $\mathbb{S}$ (i.e., $Rows \subset \beta_k$). $\beta_k$ serves as the atomic unit for transport and internally adopts a columnar memory layout. This design is critical for enabling Zero-Copy transmission between the network layer and application memory.

The SDF is designed to natively support streaming computation. Logically, it exposes an \textbf{\texttt{Iterator<Row>}} semantic to ensure the flexibility of row-level processing. Physically, however, multiple logical rows are encapsulated into $\beta_k$ to amortize network I/O overhead through batching. Once a $\beta_k$ arrives at the client, computation logic is triggered immediately without waiting for subsequent batches $\beta_{k+1}$, thereby accommodating the requirements for high-throughput streaming data processing in WAN environments.

An SDF can be mapped to either a single physical file or a list of files. A file list is mapped into a structured SDF, where file metadata is mapped to standard columns, and file content is mapped to a Binary-typed blob column. This blob column possesses an Expandable Capability, allowing unstructured binary content to be read in real-time as a new SDF. Conversely, for structured physical files (e.g., CSV, Parquet), DACP directly encapsulates their internal rows and columns into a single SDF.

\subsection{Core Methods: GET, PUT, and COOK}

To enable standardized operations on scientific data, DACP defines three fundamental methods: GET, PUT, and COOK.

\begin{itemize}

\item \textbf{GET}: 
The method is employed to request existing, static data resources. The client provides the target resource URI, and upon successful authentication, the server returns the content in the form of a Streaming Data Frame (SDF). Crucially, \texttt{GET} supports \textbf{predicate pushdown}, enabling filtering logic to be executed on the server side, thereby circumventing the movement of massive datasets across the network.

\item\textbf{PUT}:
The method is designed to upload and persist client-side data streams to the server. The input is similarly formatted as an SDF, facilitating the efficient, streaming ingestion of data into data centers.

\item \textbf{COOK}: The method serves as the gateway interface for computational task offloading. Unlike static data retrieval, the \texttt{COOK} method accepts a \textbf{Directed Acyclic Graph (DAG)} as its functional payload. 
A task is formally represented as $\mathcal{G} = (\mathcal{V}, \mathcal{E})$, where each vertex $v \in \mathcal{V}$ denotes a standardized \textit{Operator} (e.g., Filter, Select), and each edge $e \in \mathcal{E}$ represents the streaming flow of SDFs. By adopting a non-blocking mechanism, \texttt{COOK} allows clients to consume or visualize the initial batch of results in real-time.

\end{itemize}

\subsection{Resource Addressing}

DACP adopts a unified resource addressing scheme alongside a phased interaction model to facilitate the precise localization of scientific resources within heterogeneous environments. A standard URI is defined to uniquely identify an SDF across Wide Area Networks (WANs), with the naming convention specified as follows:

\begin{equation}
    \texttt{dacp://}\langle \textit{host} \rangle:\langle \textit{port} \rangle / [\langle \textit{dataset\_name} \rangle] / \langle \textit{path} \rangle
\end{equation}

\begin{itemize}
    \item \textbf{dacp}: The protocol scheme identifier.
    
    \item \textbf{host:port}: The network address and service port of the DACP server.
    
    \item \textbf{dataset\_name}: An optional dataset name. A Dataset serves as a logical collection unit for SDFs. It supports the definition of shared metadata or permission policies at the collection level, enabling all enclosed SDFs to automatically inherit this contextual information.
    
    \item \textbf{path}: The specific resource path of the SDF. This can point to a concrete physical file path or a logical dataset directory.
\end{itemize}

\begin{figure}[htbp]
    \centering  % 【推荐】使用这个命令来居中，比 \centerline 更规范
    % width=0.8\columnwidth 表示将图片宽度设置为当前栏宽的 80%
    % 你可以修改 0.8 为 0.5, 0.9, 1.0 等任意比例
    \includegraphics[width=1.0\columnwidth]{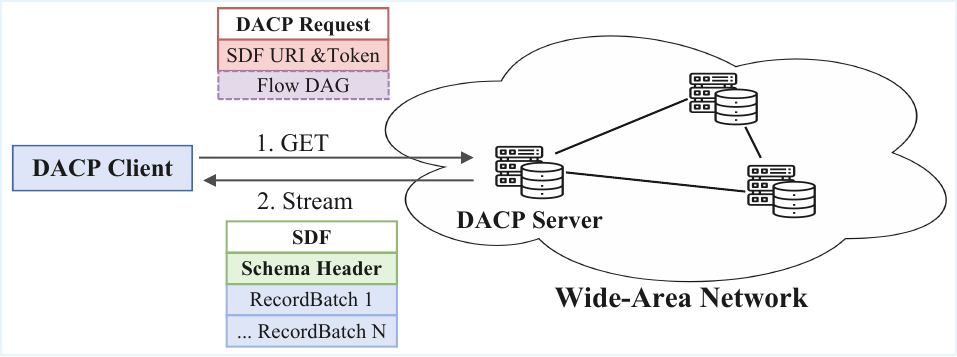}
    \caption{The Client-Server interaction mechanism of DACP. }
    \label{traditional}
\end{figure}

The DACP interaction process initiates with the client establishing a connection and exchanging credentials to obtain a short-lived access Token. Subsequently, the client issues data requests to server. For static retrieval via GET, the client consumes the SDF through a row-based iterator, strictly adhering to lazy evaluation principles to minimize latency.

\subsection{Cross-Domain Collaboration}

DACP enables cross-domain data collaboration by orchestrating the DAG defined in the \texttt{COOK} method across distributed environments.

During the scheduling phase, first DACP provides end-to-end task planning, which includes data discovery, acquisition, processing, analysis, and result integration, and maps them to multiple physical sub-tasks. Second, it ensures task execution reliability through DAG-based sub-task scheduling and transaction control. Finally, it establishes mechanisms for task scheduling, monitoring, and fault handling. Regarding security, DACP establishes a token-based data flow mechanism. Downstream nodes must present a short-lived token, issued during the scheduling phase, to \textit{pull} data from upstream nodes.

\begin{figure}[htbp]
    \centering  % 【推荐】使用这个命令来居中，比 \centerline 更规范
    % width=0.8\columnwidth 表示将图片宽度设置为当前栏宽的 80%
    % 你可以修改 0.8 为 0.5, 0.9, 1.0 等任意比例
    \includegraphics[width=1.0\columnwidth]{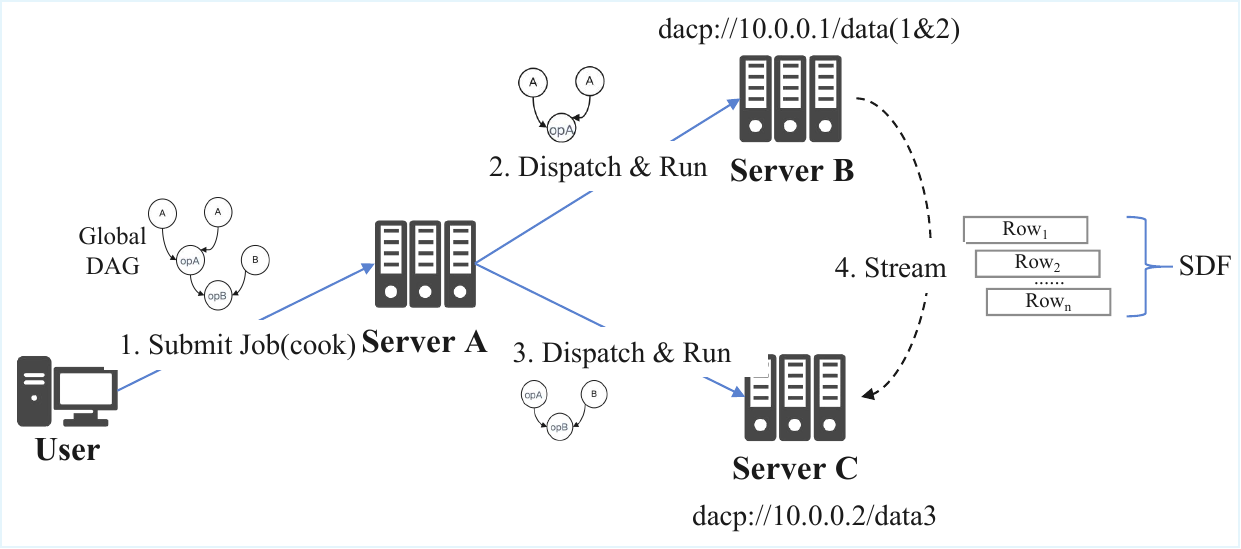}
    \caption{A cross-domain collaborative analysis scenario, where the global logical DAG is decomposed into physical sub-tasks for in-situ execution and synchronized streaming.}
    \label{fig:cross_domain_scheduling}
\end{figure}

% \begin{figure}[htbp]
%     \centering  % 【推荐】使用这个命令来居中，比 \centerline 更规范
%     % width=0.8\columnwidth 表示将图片宽度设置为当前栏宽的 80%
%     % 你可以修改 0.8 为 0.5, 0.9, 1.0 等任意比例
%     \includegraphics[width=0.9\columnwidth]{faird-arch_cropped.pdf}
%     \caption{The system architecture overview of faird.}
%     \label{fig:faird_arch}
% \end{figure}

In the task execution phase, DACP employs a lazy loading and streaming execution paradigm. Upon the client offloading a task to the server, full-scale computation is not triggered immediately. Instead, the actual computation process is initiated by the outermost node of the DAG (i.e., the result consumer), which recursively \textit{pulls} data from upstream dependency nodes; only at this point are the upstream operators activated for execution. Building on this mechanism, data is no longer loaded as monolithic files but flows between operators in units of fine-grained logical rows. This design significantly enhances the processing efficiency for large-scale scientific data.

\section{faird: The Reference Server Implementation of DACP}

This paper presents the design and implementation of faird, the reference server for the DACP protocol. In terms of architectural layering, \textbf{Apache Arrow Flight}\cite{apache_arrow} serves as the underlying \textit{Transport Layer}, responsible for providing a high-performance, zero-copy data communication channel. Faird is constructed atop this foundation as the \textit{Application Layer}. This section elaborates on the core components of faird following a bottom-up approach.

\subsection{Multimodal Data Source}

To abstract away the heterogeneity inherent in scientific data formats (e.g., CSV, NetCDF, TIFF, and arbitrary binary formats), faird implements a Multimodal Data Source framework. This abstraction is responsible for mapping specific physical files into logical SDF representations. For structured data, the source leverages memory mapping (mmap) to directly project data into the Apache Arrow format, thereby achieving zero-copy read efficiency. Conversely, for semi-structured or unstructured data, the source adheres to the ``File List Framing'' strategy described previously, effectively integrating unstructured content into the unified logical SDF view.

\subsection{SDF Engine}

The SDF Engine serves as the core kernel of faird, governing the lifecycle management and logical abstraction of in-memory SDF data. It is responsible not only for mapping physical file types to standard Arrow types but also for defining and maintaining column definitions to facilitate schema-aware computation for downstream operators. Notably, the initialization of an SDF does not trigger immediate data loading; physical data is materialized into memory as Arrow structures only during the DAG execution phase. Additionally, the engine incorporates a built-in library of fundamental operators (e.g., \texttt{Map}, \texttt{Filter}, \texttt{Select}). These operators are capable of executing high-performance batch computing directly on the Arrow columnar in-memory data layout.

\subsection{Ecosystem Integration}

To achieve seamless interoperability with mainstream computing frameworks, faird provides native adapters for Apache Spark, PyTorch, and HuggingFace. In the Spark scenario, faird functions as a high-performance data source, automatically mapping read requests to the protocol's underlying \texttt{GET} or \texttt{COOK} primitives. For AI training, faird feeds data streams directly into the PyTorch pipeline via a streaming adapter. Furthermore, it deeply integrates with HuggingFace, allowing users to load remote data directly via the \texttt{dacp://} protocol within \texttt{load\_dataset}. This capability efficiently supports the training and fine-tuning of large models.

\subsection{faird Client}

The faird client is a lightweight software development kit (SDK) implementation of the DACP protocol, designed to mask the complexity of underlying communications. It does not execute computations directly but provides a chainable API. Users can construct a logical DAG using built-in or user-defined operators. Upon triggering a computation request, the Client serializes the constructed DAG and submits it to the server. For structured data returned by server, the client leverages Apache Arrow's zero-copy mechanism to convert it directly into native in-memory objects; for unstructured Blob columns, it reconstructs them into file objects or byte streams on the fly.

\section{Performance Evaluation}

To validate the performance efficiency of the DACP protocol in real-world data transmission scenarios—specifically contrasting it with the traditional FTP protocol—we conducted a series of comparative experiments on a high-performance computing node.

\subsection{Experimental Setup}

The experiments were conducted on a high-performance computing node equipped with an Intel Xeon Gold CPU and 376GB RAM. To minimize storage bottlenecks, the system utilizes a disk array with a sequential throughput of 1202 MB/s. The network environment provides 3.45 Gbps bandwidth with negligible latency (0.2 ms).

\subsection{Datasets}
To comprehensively evaluate DACP across heterogeneous scientific data, we constructed a mixed testbed:

\begin{itemize}
    \item \textbf{Structured Data:} We utilized the Yelp Open Dataset~\cite{yelp_dataset}, selecting a subset of 1 million user review rows in JSON format. Each row maintains a uniform schema with five key-value pairs.
    
    \item \textbf{Unstructured Mixed Data:} Based on the ImageNet~\cite{deng2009imagenet} validation set, we designed a random mixed workload to simulate complex scientific repositories. This dataset consists of a random distribution of \textbf{1 large file (1GB)}, \textbf{10 medium files (100MB each)}, and \textbf{10,000 small files (10KB each)}, aiming to verify the comprehensive transmission efficiency under hybrid loads.
\end{itemize}

\begin{figure}[htbp]
    \centering  % 【推荐】使用这个命令来居中，比 \centerline 更规范
    % width=0.8\columnwidth 表示将图片宽度设置为当前栏宽的 80%
    % 你可以修改 0.8 为 0.5, 0.9, 1.0 等任意比例
    \includegraphics[width=1.0\columnwidth]{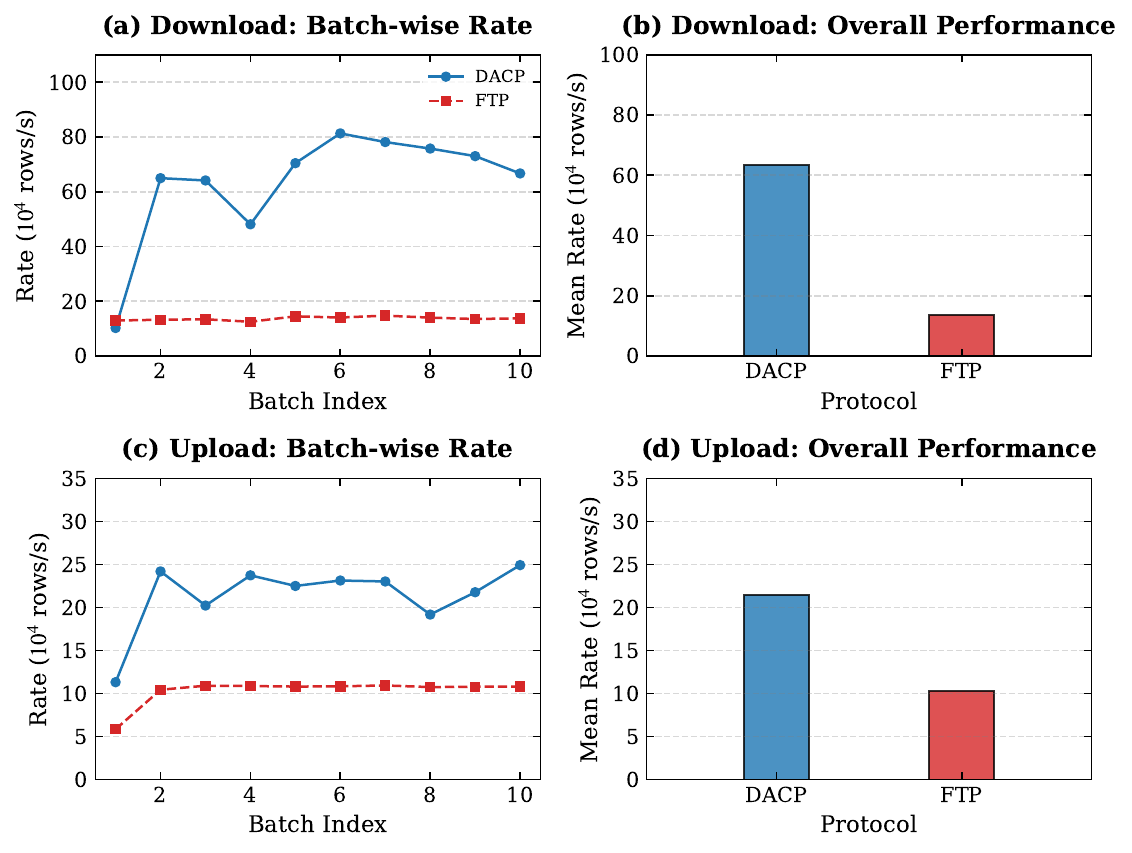}
    \caption{Performance evaluation on structured data.}
    \label{fig:eval_structured}
\end{figure}

\begin{figure}[htbp]
    \centering  % 【推荐】使用这个命令来居中，比 \centerline 更规范
    % width=0.8\columnwidth 表示将图片宽度设置为当前栏宽的 80%
    % 你可以修改 0.8 为 0.5, 0.9, 1.0 等任意比例
    \includegraphics[width=1.0\columnwidth]{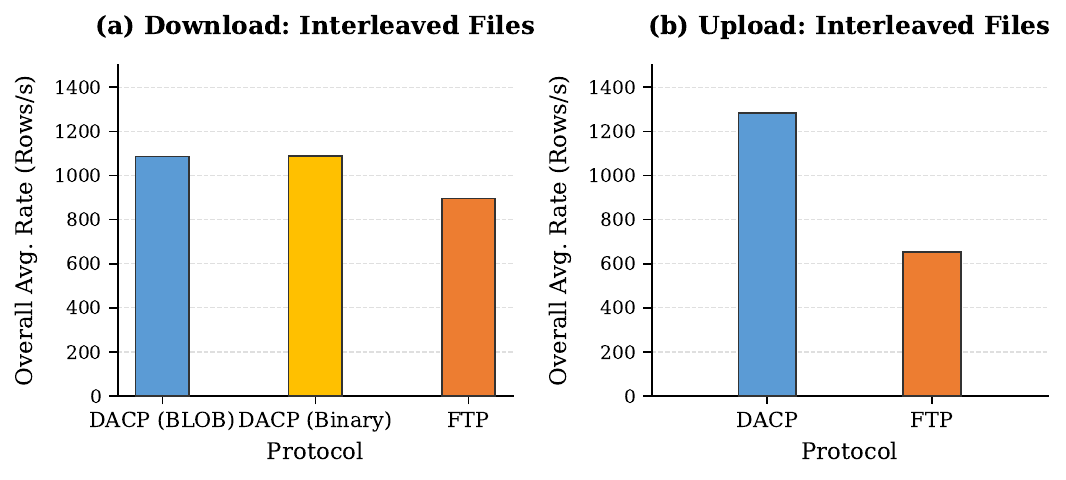}
    \caption{Performance evaluation on unstructured data(mixed).}
    \label{performance2}
\end{figure}

The experimental results demonstrate that DACP significantly outperforms FTP in overall performance across most scenarios, achieving speedup ratios ranging \textbf{from 1.21$\times$ to 5.3$\times$}. Specifically, for structured data, DACP surpasses FTP across all metrics, with speedups between \textbf{3.10$\times$ and 5.36$\times$}. For unstructured data, the performance of DACP (BLOB) is comparable to that of DACP (Binary) and slightly superior to FTP, yielding a speedup of \textbf{1.21$\times$}. Furthermore, DACP maintains symmetrical download and upload rates, whereas FTP exhibits a 13\%--27\% degradation in upload rates under random interleaved scenarios. These results indicate that DACP possesses a distinct performance advantage over FTP, particularly in complex workload scenarios.

\section{Application Scenarios}

\subsection{Efficient and Unified Data Provisioning}
Addressing the fragmentation of data across relational databases, file systems, and object stores, DACP abstracts these heterogeneous sources into a unified SDF representation. This provides a consistent access model, allowing scientists to utilize a single codebase and query interface for diverse data types without developing complex adapter layers. 

\subsection{Interactive Data Exploration}
DACP significantly optimizes interactive exploration on PB-scale datasets by mitigating the bottlenecks of traditional ``download-then-process'' paradigms. DACP allows users to submit a DAG containing filter logic, enabling the server to return only the requested slices as a Streaming Data Frame (SDF). This ``seek-and-stream'' capability achieves near real-time responsiveness, drastically improving exploration efficiency.

\subsection{Cross DataCenter Collaborative Workflows}
DACP optimizes cross-data center collaboration by replacing the costly replication of intermediate results with an in-situ computation strategy. Adhering to the principle of moving operators, not data, DACP employs lazy evaluation: upstream computation and Arrow Flight streaming are triggered only upon downstream demand. This ensures that only filtered, high-value data streams—rather than raw massive datasets—traverse the WAN.

\section{Conclusion}

In response to the limitations of legacy protocols (e.g., HTTP/FTP) in supporting data-intensive paradigms like AI4Science, this paper proposes \textbf{DACP}, a Data Access and Collaboration Protocol tailored for scientific discovery.

DACP redefines the semantics of scientific data access. By establishing the \textbf{Streaming Data Frame (SDF)} as a standard unit, it effectively abstracts the heterogeneity of underlying storage systems. Through unified resource addressing, columnar streaming, and reverse provisioning mechanisms, DACP enables cross-domain collaboration via data discovery, in-situ computation, and result streaming. We presented faird, a reference implementation that validates the feasibility of DACP in building collaborative infrastructures, with experiments demonstrating its superior performance across various scenarios.

In the future, DACP aims to empower scientists to access and analyze remote data as transparently and efficiently as local resources, thereby fully unleashing the potential of data elements in scientific discovery.

\section*{Acknowledgment}

This work is supported by the National Key R\&D Program of China (Grant No. 2021YFF0704200), and the Informatization Plan of Chinese Academy of Sciences (Grant No. CAS-WX2022GC-02).

\bibliographystyle{IEEEtran}
\bibliography{ref}

\end{document}